\documentstyle{mn}
\input psfig.sty
  
\def\solmas{{M$_\odot$}}
\def\simless{\mathbin{\lower 3pt\hbox
   {$\rlap{\raise 5pt\hbox{$\char'074$}}\mathchar"7218$}}}   
\def\simgreat{\mathbin{\lower 3pt\hbox
   {$\rlap{\raise 5pt\hbox{$\char'076$}}\mathchar"7218$}}}   
\def\etal{{\rm et al.}}

\def\solmas{{M$_\odot$}}

\def\mnras{{MNRAS}}

\def\be{\begin{equation}}
\def\ee{\end{equation}}

  \makeatletter
  \ifx\CUP@mtlplain@loaded\undefined  
  \makeatother  
    
  \else
    
  \fi

\loadboldmathitalic 
\loadboldgreek      
  
\makeatletter
\ifx\CUP@mtlplain@loaded\undefined  
  \newfont\bit{cmbxti10 at 9pt}
\else
  \newfont\bit{mtbxti10 at 9pt}
\fi
\makeatother

\def\LaTeX{L\kern-.36em\raise.3ex\hbox{a}\kern-.15em
    T\kern-.1667em\lower.7ex\hbox{E}\kern-.125emX}

\newcommand{\gsim}{\mathrel{\hbox{\rlap{\lower.55ex \hbox {$\sim$}}
                   \kern-.3em \raise.4ex \hbox{$>$}}}}
\newcommand{\lsim}{\mathrel{\hbox{\rlap{\lower.55ex \hbox {$\sim$}}
                   \kern-.3em \raise.4ex \hbox{$<$}}}}

\title[] {Star Formation in Transient Molecular Clouds}
\author[Clark \&  Bonnell]
{Paul C. Clark\thanks{E-mail: pcc@st-and.ac.uk} and Ian A. Bonnell\\
School of Physics Astronomy, University of St Andrews, North Haugh, 
St Andrews, Fife, KY16 9SS. \\ }

\date{\today}

\begin{document}

\maketitle

\begin{abstract}

We present the results of a numerical simulation in which star formation
proceeds from an initially unbound molecular cloud core. The turbulent
motions, which dominate the dynamics, dissipate in shocks leaving a
quiescent region which becomes gravitationally bound and collapses to form a
small multiple system. Meanwhile, the bulk of the cloud escapes due to its
initial supersonic velocities. In this simulation, the process naturally
results in a star formation efficiency of $\sim 50$\%. The mass involved in
star formation depends on the gas fraction that dissipates sufficient
kinetic energy in shocks. Thus, clouds with larger turbulent motions will
result in lower star formation efficiencies. This implies that globally
unbound, and therefore transient giant molecular clouds (GMCs), can account
for the low efficiency of star formation observed in our Galaxy without
recourse to magnetic fields or feedback processes. Observations of the
dynamic stability in molecular regions suggest that GMCs may not be
self-gravitating, supporting the ideas presented in this letter. 

\end{abstract}


\section{The star forming environment}

The formation of stars in giant molecular clouds appears to be a very
inefficient process. Global estimates of the star formation efficiency, based on the mass in molecular clouds
and the Galactic star formation rate(Scalo 1986; Evans 1999), are of order a
few percent while small-scale estimates for stellar clusters approach
50\% (Lada 1992; Lada \& Lada 2003).  The dispersion in estimates can, to
some extent, be decreased under the assumption that molecular clouds are
long-lived entities that exist for many dynamical times, although this necessitates a supporting mechanism to balance gravity (McKee \etal 1993). 

As star formation from molecular gas involves the gravitational contraction
through many orders in magnitude in size, it is commonly assumed that the
largest scale objects associated with star formation, molecular clouds, are
themselves bound. In this case, the problem is how does a large bound region
($10^{4} - 10^{6}$ \solmas ) only permit a small fraction of its mass to
undergo gravitational collapse. Various mechanisms, such as magnetic fields
and feedback from young stars, have been evoked over the years in order to
support molecular clouds and explain the low efficiency of star formation
(Shu, Adams \& Lizano 1987; Franco, Shore \& Tenorio-Tagle 1994). In this
letter, we explore an alternative scenario whereby molecular clouds are not
bound on the large scale, circumventing the need for any additional
supporting mechanisms.

In a recent paper, Elmegreen (2000) has collected observational evidence
implying that GMCs are short-lived objects, with dispersal times comparable
to their crossing times ($t_{dyn} \sim 3 \times 10^{7}$years). The formation of
stars would then have to occur in a short burst ($\sim t_{dyn}$), as has been
suggested from recent observations (Hartmann, Ballesteros-Paredes, \& Bergin
2001; Hartmann 2002), rather than an ongoing process. This implies that molecular
cloud formation is dynamical, and self-gravity need not play a
dominant role. Such a scenario has recently been advanced by Pringle, Allen
\& Lubow (2001) where the passage of a spiral arm triggers the agglomeration
of unbound molecular gas and the ensuing shock dissipates sufficient kinetic
energy to allow self-gravity to (locally) form stars. The observed molecular
clouds could then be transient structures with lifetimes comparable to their
dynamical times (Elmegreen 2000; Pringle et al 2001; Larson
2003). In this picture it would seem likely that these regions are globally
unbound and highly dynamic, with only a small fraction achieving
gravitational instability before the cloud dissipates. This is possible if
the internal bulk motions of the cloud are stronger than the
self-gravity of the region. The supersonic nature of the motions
produces shocks which dissipate kinetic energy, eventually leading to
regions becoming gravitationally unstable. The efficiency of the star
formation would then be governed by the strength of the internal kinetic
energy at the point of the cloud's formation, along with how much support
can be removed in shocks. This would naturally lead to different clouds
having different star formation efficiency and potentially explain the
variance in the observed estimates.

Observational estimates of the mass, and therefore the energies, of the molecular gas is inherently difficult and uncertain.
Heyer, Carpenter \& Snell (2001) have investigated the dynamic stability of
molecular regions in the outer Galaxy. Results suggest that the molecular
material is unbound until the mass of the region approaches $10^{5}$\solmas.
It is however noted in the paper that obtaining a value for the hydrogen
mass via measurements of CO flux involves the assumption that the gas is
virialised. If the region in question is then unbound, the subsequently
derived value for the mass of molecular hydrogen is an over-estimate of the
region's true mass. As a result, regions will appear more bound than
they actually are. This places serious doubt on the generally accepted idea
of globally bound GMCs.

The ability of turbulence to character the interstellar medium (ISM) along with its
role in star formation in GMCs has been extensively studied by a number of
authors. For a complete review we point the reader to the article by
Mac Low \& Klessen (2003). It is now well established that large scale turbulent
motions can produce the filamentary structure that is commonly observed in star
forming regions (V{\'a}zquez-Semadeni, Ballesteros-Paredes \& Rodriguez 1997;
Ostriker, Gammie \& Stone 1998; Ballesteros-Paredes, Hartmann \& V{\'a}zquez-Semadeni
1999; Padoan \& Nordlund 1999) and that even MHD turbulence decays on a dynamical timescale (Mac Low et al
1998; Stone, Ostriker \& Gammie 1998) such that a driving mechanism is required if
the clouds are to exist for many dynamical times. These studies highlight that
transient objects can form from large scale flows in the ISM, supporting the work
of Elmegreen (2000) that star forming regions are short lived in nature. Recent
work on subregions within GMCs has also shown that driven turbulence, and more
importantly the scale of the driving, can effect the local efficiency of the star
formation (Klessen, Heitsch \& Mac Low 2000; Klessen 2001). The periodic boundary
conditions in this last study make it difficult to find estimates of the global
star formation efficiency of transient molecular regions, since the mean mass
density cannot decrease. Any newly formed stars can thus continue to accrete mass
indefinitely. Global calculations have tended to study clouds that are only
marginally unbound and hence do not provide information on transient GMCs (Bate,
Bonnell \& Bromm 2003; Bonnell, Bate \& Vine 2003)

It is the picture of GMC formation presented by Pringle et al (2001) that we
use as the motivation for this letter. They suggest that molecular clouds
are formed via the accumulation of small parcles of material. In this
manner, we interpret the internal motions of the cloud as a result of the
original random motions of the constituent parcles of gas. The internal
motions therefore need not be turbulent in the strict definition and so we
use the notion of ``turbulence'' loosely in this paper. The formation mechanism
for molecular regions supported here is distinctly different from that
proposed by Ballesteros-Paredes et al (1999). Since we are concerned with a
transient object and one which is not part of some large scale flow, we do
not require boundary conditions on our model. Futhermore, since the object
in study is transient, we do not include any feedback effects into the
internal motions as is consistent with the ideas presented by Elmegreen
(2000).

We present here the first numerical simulation in which the turbulent gas is
both initially unbound and free to expand, showing that this does result in
star formation but with a relatively low efficiency. In section 2 we
describe the details of the simulation technique with a discussion of the
evolution and efficiency of the star formation given in section 3. We
summarise and discuss the implications of this result in section 4.

\begin{figure}
\centerline{\psfig{figure=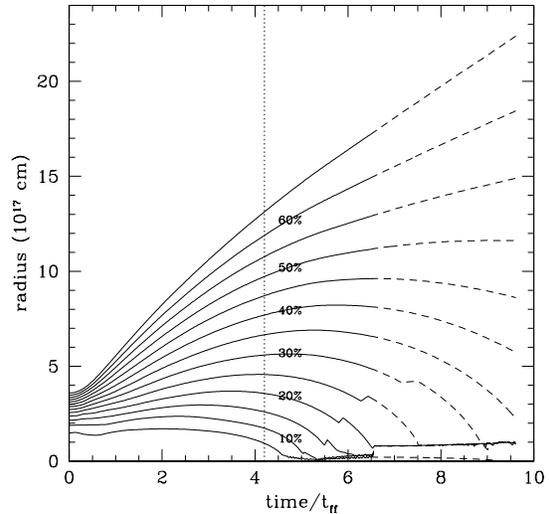,width=2.9truein,height=2.9truein}}
\caption{\label{lagradii} The evolution of Lagrangian radii enclosing a fixed fraction of the total mass. The solid lines are from the original simulation and the dashed lines are from the period where the smallest binaries are merged. The vertical dotted line denotes the time in the simulation at which the first protostar is formed.}
\end{figure}

\begin{figure*}
\centerline{\psfig{figure=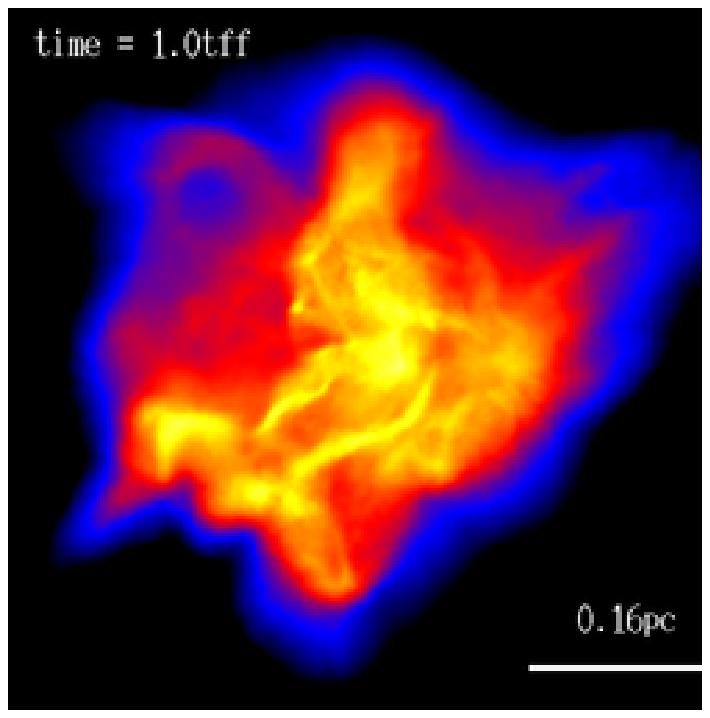,width=1.8truein,height=1.8truein}
\psfig{figure=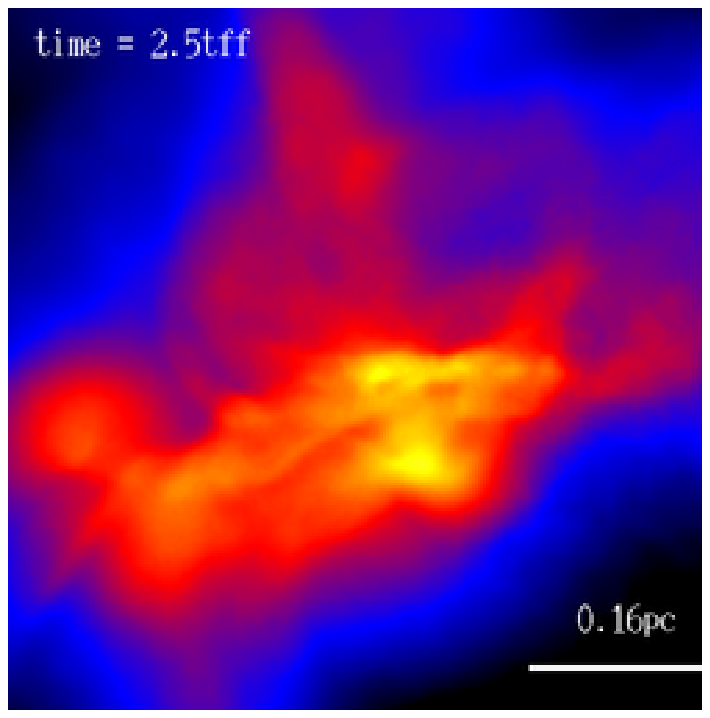,width=1.8truein,height=1.8truein}
\psfig{figure=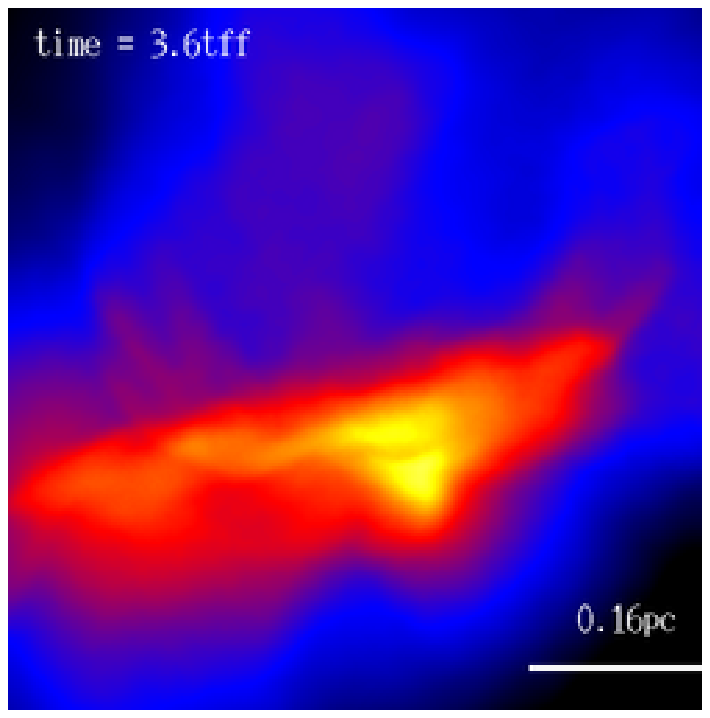,width=1.8truein,height=1.8truein}}
\centerline{\psfig{figure=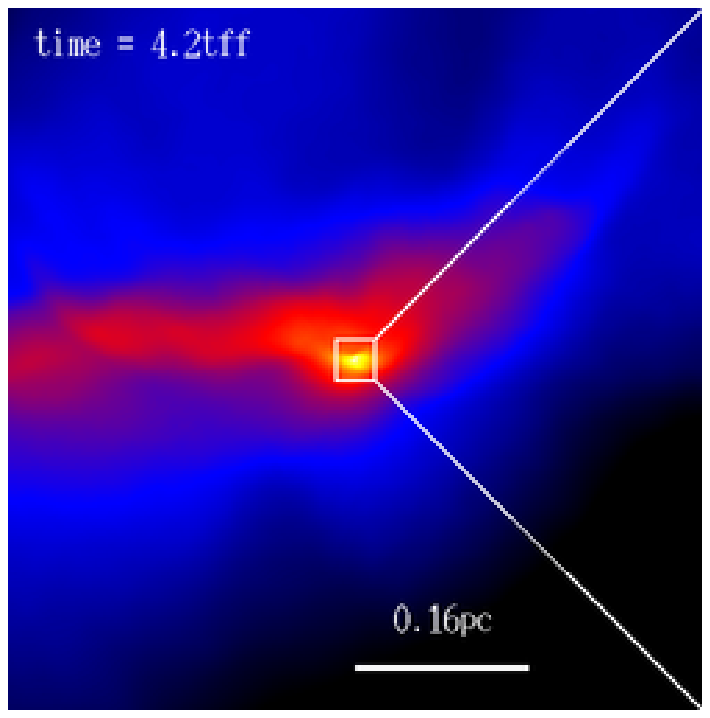,width=1.8truein,height=1.8truein}
\psfig{figure=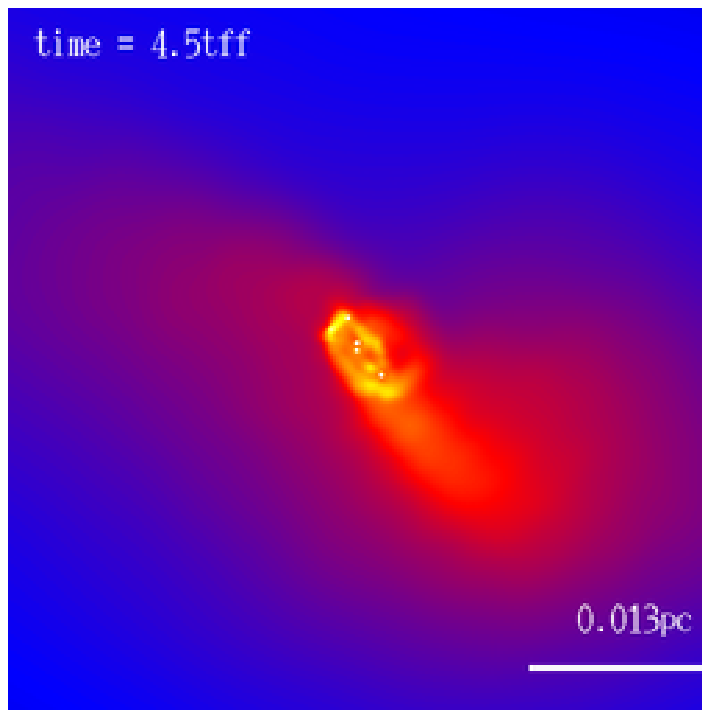,width=1.8truein,height=1.8truein}
\psfig{figure=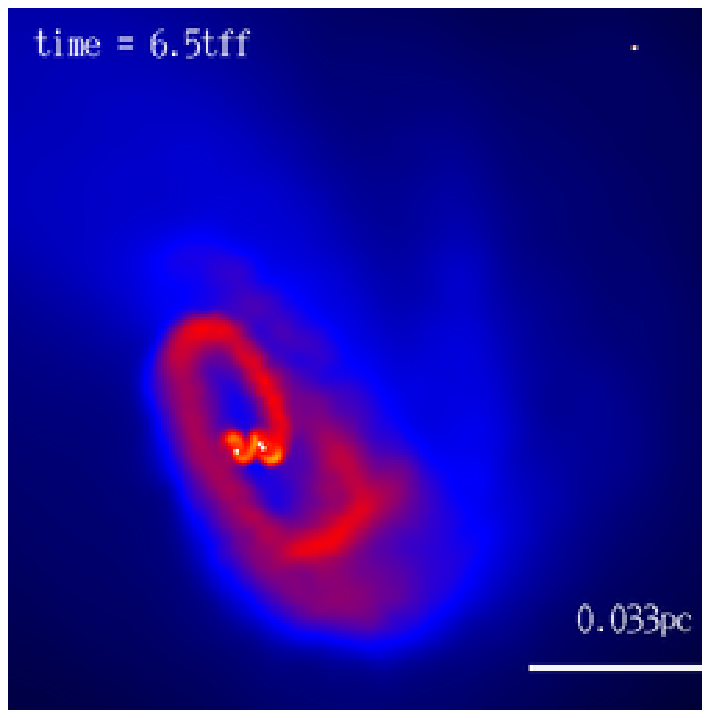,width=1.8truein,height=1.8truein}}
\caption{\label{evolpics} The panels show logarithmic column density snapshots from the simulation, following the evolution of the molecular cloud. The maximum densities in the first four panels are 0.21, 0.17, 0.13 \& 3.8 ${\rm{g\;cm^{-2}}}$ respectively and the last two panels have a maximum density 100${\rm{g\;cm^{-2}}}$, taken to arbitrarily represent the protostars. Minimum column densities are 0.001${\rm{g\;cm^{-2}}}$ for the first four panels and 0.01${\rm{g\;cm^{-2}}}$ for the last two. The turbulent velocities injected into the cloud at the beginning of the simulation quickly erases the uniform density and imposes a characterising filamentary structure. The kinetic energy is larger than the potential causing the cloud to expand, with the outer layers escaping the self-gravity of the system. Left behind by the escaping mass is a small dense region. The gas at the centre of the dense region, it is able to rid itself of kinetic energy via shocks with the shrouding material. Eventually gravity manages to dominate and produce a dense core $(t \sim 4t_{ff})$ which fragments to form a 3 body hierarchical system with one protostar being ejected (visible at the top right of the last panel in the figure).}
\end{figure*}

\section{ The simulation }

The simulation starts from a uniform sphere of mass 31.6\solmas$\;$at 10K
with a radius of 0.13pc. The material is modelled as an ideal gas of
molecular hydrogen and the equation of state is isothermal. With these
properties, the Jeans mass of the cloud is 1\solmas$\;$and the Jeans number,
$J_{o} = (M/M_{Jeans})^{3/2} = E_{grav}/E_{thermal} = 10$. These conditions give an initial density of $2.2 \times
10^{-19} {\rm{g\;cm^{-3}}}$ and hence the free-fall time, $t_{ff}$, is $1.4
\times 10^{5}$ years. The gas is given turbulent support modelled by a
Gaussian random field with a power spectrum $P(k) \propto k^{-4}$, where $k$
is the wavenumber of the velocity perturbations and corresponds to a Larson
type law of $\Delta v \propto L^{-0.5}$ (Larson 1981; Ostriker, Stone \&
Gammie 2001). The turbulent kinetic energy of this study is characterised by
the parameter $\epsilon = E_{grav}/E_{kinetic}$ and it is the unbound case
of $\epsilon = 0.75$ which is presented here, resulting in a turbulent Mach
number (as calculated from the kinetic energy) of $\sim 5.5 \;(c_{s} \approx
0.2{\rm{km s^{-1}}})$. 

The fluid was modelled using the Lagrangian method of smoothed particle
hydrodynamics (SPH). Gravitational forces are calculated using a tree
structure (Benz et at 1990). The code includes the modification by Bate et
al (1995) which replaces dense bound regions of the gas with point masses,
or `sink particles'. These sinks allow the code to model the dynamical
evolution of accreting protostars, without integration times becoming
prohibitively small. Sink-particle creation occurs when the densest gas
particle (at a given time) and its 50-100 neighbours are self-gravitating,
subvirial and occupy a region smaller than the sink-radius. In this
calculation we used a sink radius of 48AU with the sinks being formed at a
density of ~$1 \times 10^{-14} {\rm{g\;cm^{-3}}}$. To minimise computational
expense, we smooth the gravitational forces between stars at a distance of
33AU. We use 204800 SPH particles to model the gas and are hence able to
resolve self-gravitating objects as small as 0.015\solmas$\;$(Bate \&
Burkert 1997). 

At the end of the simulation 4 protostars were formed containing roughly
17\% of the mass of the original cloud. The formation of a tight binary with
a  separation of 134AU, considerably reduced the time step of the
integration making it computationally expensive to run the simulation for
longer than $6.5t_{ff}$. Unfortunately, this prevented satisfactory
examination of the outer regions in the cloud, which evolve on much longer
time-scales than the material near the stellar system. Replacing the binary
with a point mass, we were able to run the simulation much further, allowing
a better examination of the evolution of the outer regions. The results of
this part of the simulation are represented by the dashed radii curves in
figure {\ref{lagradii}}. It must be noted here that for a radius of less
that $~3 \times 10^{17}$cm from the protostars the results for the accretion
are not as accurate as the original simulation due to the large accretion
radius of the point mass, but this does not affect the results outside this
region.  All calculations of kinetic energies and radii in this letter are
obtained from the centre of mass of material that will eventually end up in
the stars, rather than the centre of mass of the entire cloud. The
Lagrangian particle nature of SPH makes this possible. This gives a picture
of the star formation process as seen by the material that is actually
involved and allows for a better estimate of the star formation efficiency.

\section{Evolution}

\subsection{General Properties}

The evolution of this cloud can be split roughly into three phases:
turbulent compression, general expansion of the gas, and a final
gravitational collapse for a subset of the material. Column density
snapshots from the simulation can be seen in figure {\ref{evolpics}},
showing the effect of the turbulent motions and how the gas eventually
evolves to form a small stellar system.

The simulation starts with a uniform sphere of gas with a turbulent energy
field. The large random motions quickly dominate with shocks producing the
filamentary structure typical of turbulent regions. Densities at the centre
of these filaments reach $\sim 25$ times the initial density of the cloud.
Despite this increase in density, they are not self-gravitating structures.
In fact they are highly transitory, existing for only a few tenths of a
free-fall time. The turbulent motions lead to essentially one dimensional
compression which does not produce a decrease in the local Jeans mass
(Doroshkevich 1980; Lubow \& Pringle 1993; Clarke 1999) and hence the gas is
generally unbound. A 1-D compression does not decrease the effective
gravitational radius of an unbound parcel of gas. Thus, it cannot alter the
boundness of a clump unless it decreases its thermal energy content.
Significantly, these transitory structures provide the seeds for the
fragmentation that occurs during the later collapse phase.

Figure {\ref{lagradii}} shows the Lagrangian radii that contain fixed
fractions of the cloud's mass, and how these change with time. The global
expansion is readily apparent from the figure, as is the extent of the
expansion by the onset of star formation. The initial turbulence decreases
during the first free-fall time, leading to a general expansion of the
cloud. After the random motions have been damped out in the shocks,
expansion is the main component of the velocities. During this phase, most
of the detailed filamentary structure is erased, with only one large
elongated body of gas remaining.

Around $t = 2t_{ff}$, the expansion for a small fraction of the material, up
to roughly 5\%, stops and this gas starts to fall back in. As the Lagrangian
radii show, this is an inside out process. The inner 5\% starts contracting
before the outer layers and in-fall occurs progressively later as we move
away from the centre. It does differ however from the classic inside-out
process described by Shu (1977), since the system is not collapsing from a
static starting point, but rather from the re-contraction of an initially
expanding medium. The inner 5\% corresponds roughly to a thermal Jeans mass
and is able to collapse once the kinetic energy has been dissipated.
Gravitational contraction forms a small dense core which becomes the natal
site for a stellar system. The highly aspherical form of the core allows
fragmentation to occur, preventing the formation of a single star. The gas
splits into four protostars, arranging themselves in two tight binaries.
This system then decays with one star being ejected and the remaining 3 left
in a hierarchical triple system. At the end of the initial simulation, the
protostars contain roughly 17\% of the original mass of the system, and a
further 5\% is incorporated into the various circumbinary/stellar discs. The
inner binary is composed of protostars of mass 0.95 and 1.79\solmas. The
third protostar that makes up the bound triple system is 2.13\solmas$\;$and
the escaper is the lightest member at 0.37\solmas.

\subsection{Star formation efficiency}

The most significant feature of this simulation is that a large fraction of
the gas is expelled from the cloud, resulting in a naturally low star
formation efficiency. Escaping gas arises due to excess and undamped energy
in the cloud. The mass that is bound therefore represents an upper limit to
the possible final system mass. In principle, the most reliable method of
determining the star forming efficiency of the cloud is to let the
simulation run until all gas has either escaped or been accreted. Obviously
this is unpractical with current computational methods, so we must look to
other methods to obtain our results. Unfortunately, simply calculating the
energy of each SPH particle at the end of simulation does not unambiguously
yield a value for the efficiency. The long-range nature of gravity is such
that unbound material contributes to the potential energy of the entire
cloud. Consequently, material that appears to be bound might actually become
unbound once the escaping regions have blown away. Fortunately, the fraction
of unbound material, and thus escaping mass, yields an accurate upper limit
to the amount of material available for star formation. 

\begin{figure}
\centerline{\psfig{figure=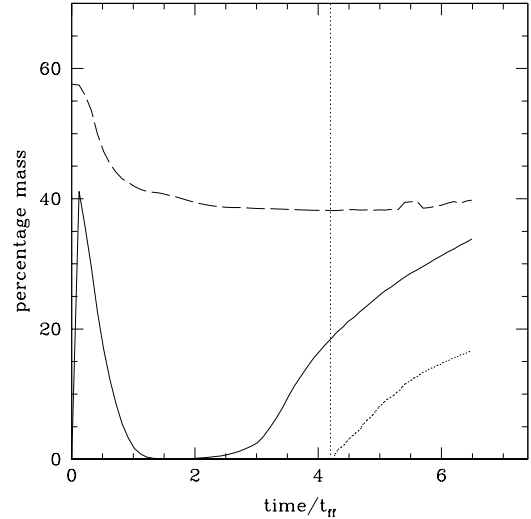,width=2.9truein,height=2.9truein}}
\caption{\label{massfrac} Shown here is the evolution of various important
features of the cloud. The long dashed line shows the fraction of material
that is unbound to the system, and the short dashed line is the amount of
material that has been accreted by the protostars. The solid line represents
the amount of material which has an increasing $|E_{grav}|$ and is
thereby involved in compression. Again, the vertical dotted line represents
the onset of protostar formation. } \end{figure}

The evolution of the unbound fraction is followed in figure
{\ref{massfrac}}. We see a sharp fall in this fraction during the first
free-fall time. This is simply a result of the decaying turbulent energy
field, as kinetic energy is dissipated in shocks. After $t=t_{ff}$, the
unbound fraction remains roughly constant for the remainder of the
simulation. At the point of star formation, at least ~40\% of the cloud is
unbound, thereby placing an upper limit of the star formation efficiency at
60\%.

The Lagrangian radii in figure {\ref{lagradii}} give another estimate for
the star formation efficiency. The dashed lines, obtained from the period in
the simulation where the close binary is merged into a single protostar,
reveal that radius containing 50\% of the mass appears to have reached a
constant value. This indicates that the 50\% fraction is the boundary
between in-falling and expanding matter. This suggests a slightly lower
value for the efficiency than we obtained by looking at the energy. In fact
the Lagrangian radii containing 60\% of the cloud shows no sign of
recontracting.

Although turbulence does not directly induce fragmentation in the cloud for
the reasons given in the previous section, it does play an important role in
determining the mass that gets converted into stars. The fraction of mass
with a decreasing potential energy represents the material that is actively
involved in compression. From figure {\ref{massfrac}} we see that this
fraction undergoes a sharp initial rise and a gentler fall during the first
free-fall time. Again this is just a consequence of the turbulent
compression and creation of transient filamentary structure. During this
phase, just over 40\% of the mass is involved in these shocks and as a
result loses kinetic energy. Having lost this energy, the gas is able to
contribute to star formation. At the end of the simulation all the mass
contained in the protostars and the circumbinary disc had originally been
involved in shocks during the turbulent phase. The turbulent compression in
this simulation is therefore linked to the fraction of the mass which is
involved in the star formation process. 

It must be noted here that in reality the tenuous outer layers of the cloud, are
unlikely to be able to find their way onto the stellar system. Clouds like the one
presented here, or cores as they are known at this scale, do not exist in
isolation but commonly in crowded regions (Motte, Andr{\'e}, Neri, 1998; Testi \&
Sargent 1998). In such an environment, tidal forces can have an important effect
on the dynamics of gas, stripping away loosely bound material. As well as
contending with tidal effects the low density regions outside dense cores are
associated with high velocity flows (Larson 1981; Padoan et al 2001), providing a
hostile environment for quiescent, tenuous gas.

\section{Implications for the star forming process}

The results presented here follow the evolution of a molecular cloud 
containing a supersonic turbulent velocity field which has greater energy
than that of the cloud's self-gravity. Despite the globally unbound nature
of the cloud, the freely decaying turbulence (from shocks) results in
dynamically quiescent regions where gravity is able evoke collapse and
subsequent star formation. Since not all of the cloud is involved in the
shocks, less than half the mass is able to undergo star formation, with the
remaining material escaping the self-gravity of the system. Four protostars
are produced by the gravitational fragmentation, three of which are in a
triple system which is formed by the ejection of the fourth body.

The fact that bound structure formed from a dynamically unbound cloud, has
important implications for the star formation process. This illustrates that GMCs
need not be globally bound in order to produce their observed stellar populations,
although more detailed calculations at the correct scale are needed to confirm
this. Internal turbulent motions are able to dissipate enough kinetic energy in
supersonic shocks to leave coherent regions where gravitational forces can
dominate. The rest of the cloud can then escape, allowing GMCs to dissolve of
their own accord instead of requiring some unbinding mechanism. The end result is
a GMC that is naturally inefficient at forming stars. This supports a large scale
picture of star formation in which GMCs are formed dynamically from the
accumulation of molecular gas (Pringle et al 2001; Elmegreen 2001). Self-gravity
need play no role in the formation and evolution of GMCs upto the point of star
formation.

Star formation from globally unbound giant molecular clouds is able to
provide a simple and interesting alternative to the efficiency problem that
is normally attributed to a combination of magnetic fields and feedback
processes. All gas which is able to dissipate enough kinetic energy to
become bound will be involved in the star forming process. The ruling factor
on the efficiency is therefore the amount of gas that is involved in shocks,
which is in turn linked to the strength  and power spectrum of the turbulent
field (Klessen et al 2000). More simulations will be needed to investigate how the efficiency is linked to both the initial Jeans number and the Mach number of the turbulence.

\section*{Acknowledgements} We would like to thank Jim Pringle, Bruce
Elmegreen, Richard Larson, Ken Rice and Steve Vine for helpful discussion and comments
which greatly improved the paper.

\end{document}